\documentclass[doublecol]{epl2}
\usepackage{color}
\usepackage{amsmath}
\usepackage{amssymb}
\usepackage{graphicx}
\usepackage{bm}

\usepackage{caption}
\usepackage{subcaption}

\title{Sub-Doppler laser cooling of fermionic $^{40}$K atoms
in three-dimensional gray optical molasses}
\shorttitle{Sub-Doppler laser cooling of fermionic $^{40}$K atoms} 

\author{D. Rio Fernandes\inst{1} \thanks{These authors contributed equally to this work.} \thanks{Email: \email{diogo.fernandes@lkb.ens.fr}}  \and F. Sievers\inst{1} $^{\rm (a)}$ \thanks{Email: \email{franz.sievers@lkb.ens.fr}} \and N. Kretzschmar\inst{1} \and S. Wu\inst{2} \and C. Salomon\inst{1} \and F. Chevy\inst{1}}

\shortauthor{D. Rio Fernandes \etal}

\institute{
  \inst{1} Laboratoire Kastler-Brossel, \'Ecole Normale Sup\'erieure, CNRS
and UPMC - 24 rue Lhomond, 75005 Paris, France\\
  \inst{2} Department of Physics, College of Science, Swansea University, Swansea, SA2 8PP.
United Kingdom
}

\pacs{37.10.De}{Atom cooling methods}
\pacs{37.10.Gh}{Atom traps and guides}
\pacs{67.85.-d}{Ultracold gases, trapped gases}

\abstract{
We demonstrate sub-Doppler cooling of $^{40}$K on the D$_{1}$
atomic transition. Using a gray molasses scheme, we efficiently cool a compressed cloud of
$6.5\times10^{8}$ atoms from $\sim 4\,\text{mK}$ to $20\,\text{\ensuremath{\mu}K}$
in $8\,\text{ms}$. After transfer in a quadrupole magnetic trap, we measure a phase space density of
$\sim10^{-5}$.  This technique offers a promising route for fast evaporation of fermionic $^{40}$K.}

\def\be{\begin{equation}}
\def\ee{\end{equation}}

\begin{document}

\maketitle

\section{Introduction}

Cooling  of fermionic atomic species
has played a fundamental role in the study of strongly correlated
Fermi gases, notably through the experimental exploration of the BCS-BEC crossover,
the observation of the Clogston-Chandrasekhar limit to superfluidity, the observation of the Mott-insulator transition in optical lattices,
and the study
of low dimensional systems (see for instance~\cite{Inguscio2006,Bloch2008} for a review). When
 the temperature is further decreased, new exotic
phases are predicted ({\it p-wave} superfluids for spin imbalanced gases, antiferromagnetic order..) and, as a consequence, intense experimental effort
is currently under way to push the temperature limit achieved in ultracold
fermionic samples in order to enter these novel regimes.

Most experiments on quantum degenerate gases begin with a laser cooling phase that is followed by evaporative cooling in a non-dissipative trap. The final quantum degeneracy strongly depends on the collision rate at the end of the laser cooling phase and sub-Doppler cooling~\cite{Dalibard1989} is often a key ingredient for initiating efficient evaporation. In the case of fermionic lithium-6 and potassium-40, the narrow hyperfine structure of the $P_{3/2}$ excited level does not allow for efficient Sisyphus sub-Doppler cooling to the red of a $F\to F'=F+1 $ atomic transition~\cite{Lin1991,Modugno1999}.

Experiments for producing quantum degenerate gases of $^{40}$K typically start with $\sim 10^8$ atoms laser-cooled to the Doppler limit ($145\,\mu$K)~\cite{DeMarco1999_2}. More refined laser-cooling schemes have produced $^{40}$K temperatures of $\sim 15\,\mu$K, but with only reduced atom numbers ($\sim 10^7$)~\cite{Modugno1999,Taglieber2008,Gokhroo2011}.
This relatively poor efficiency is due to the combination of
the fairly narrow and inverted hyperfine level structure of the $P_{3/2}$ excited state which results
in the washing out of the capture velocity of the molasses when the laser
detuning is increased~\cite{Landini2011}. To overcome
these limitations, two groups recently realized Magneto-Optical Traps
(MOT) in the near-UV and blue regions of the spectrum to cool $^{6}$Li~\cite{Duarte2011} and $^{40}$K~\cite{McKay2011} respectively. The associated transitions, being narrower than their D$_2$ counterparts, lead to a smaller Doppler temperature and were used to improve the final phase space density typically by one order of magnitude.

In this Letter, we report efficient sub-Doppler cooling of $^{40}$K
atoms using gray molasses on the D$_{1}$ atomic transition at 770~nm.
Thanks to the much reduced fluorescence rate compared to standard bright sub-Doppler molasses, we could produce cold and dense atomic samples.
  The temperature
of a tightly compressed cloud of $6.5\times10^{8}$ atoms was decreased
from $\sim4\,\text{mK}$ to $20\,\text{\ensuremath{\mu}K}$ in $8\,\text{ms}$ without significant
change of the density in the process. After transfer to a quadrupole magnetic trap, we achieved a phase space density of $\sim 2\times 10^{-5}$.

\section{Gray molasses}
\begin{figure}
\centerline{\includegraphics[width=0.8 \columnwidth]{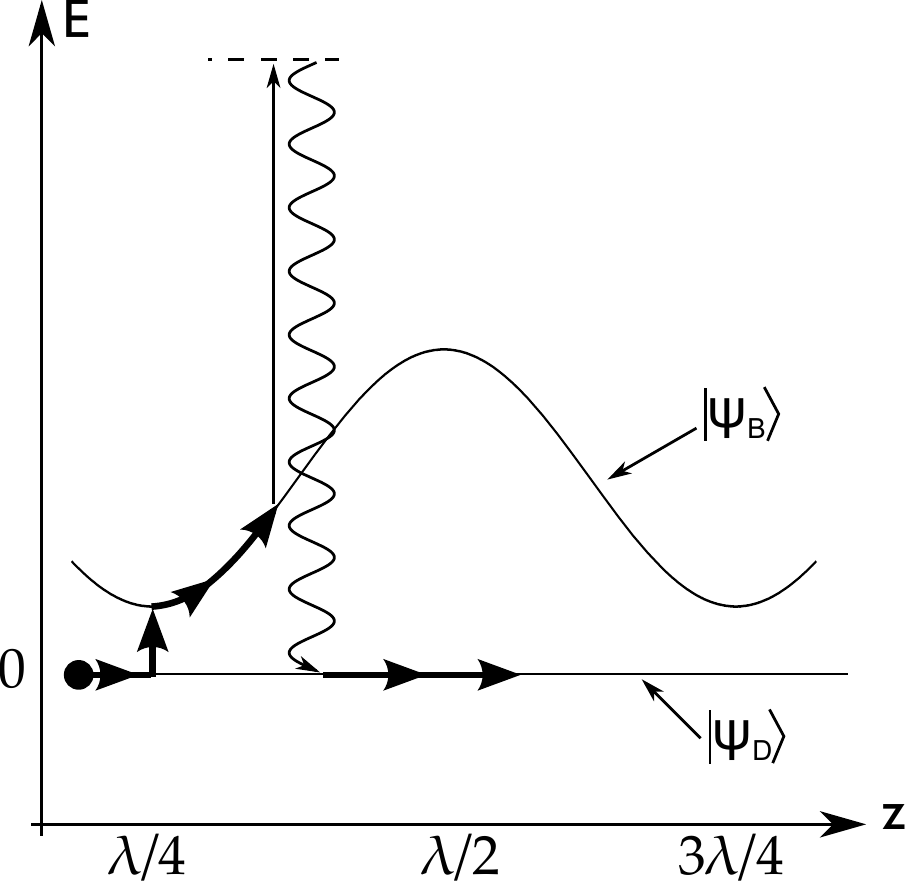}}
\caption{{\em Gray Molasses scheme.} On a $F\rightarrow F'=F$ or $F\rightarrow F'=F-1$ optical transition with positive detuning, the ground state splits into a dark and a bright manifold with positive energy, shown as $| \psi_{\rm{D}}\rangle $ and $| \psi_{\rm{B}}\rangle $ respectively. In the presence of a polarization gradient, the bright state energy is spatially modulated. Like in Sisyphus cooling, energy is lost when an atom in $| \psi_{\rm{B}}\rangle $ climbs a potential hill before being pumped back into the dark state $| \psi_{\rm{D}}\rangle $. Motional coupling between $| \psi_{\rm{D}}\rangle $ and $| \psi_{\rm{B}}\rangle $ occurs preferentially at the potential minima.}
\label{Fig:Scheme}
\end{figure}

Sub-Doppler cooling using gray molasses was  proposed in ref.~\cite{Grynberg1994} and realized in the mid '90s on the D$_2$ atomic transition of cesium and rubidium, allowing one to cool atomic samples close to 6 times the single photon recoil energy~\cite{Boiron1995,Esslinger1996,Boiron1996}.
For an atomic ground state with angular momentum $F$, gray molasses operate on the $F\to F'=F$ ($F\to F'=F-1$) optical transition. For any polarization of the local electromagnetic field, the ground state manifold possesses one (two) dark states which are not optically coupled to the excited state by the incident light~\cite{Olshanii1992,Grynberg1994}. When the laser is detuned to the blue side of the resonance, the ground state manifold splits into dark states which are not affected by light and bright states which are light-shifted to positive energy by an amount which depends on the actual polarization and intensity of the laser field (see fig.~\ref{Fig:Scheme}).

When the atom is in a bright state, it climbs up the hill of the optical potential before being pumped back to the dark state near the top of the hill. The kinetic energy of the atom is thus reduced by an amount of the order of the height of the optical potential barrier. The cooling cycle is completed near the potential minima by a combination of motional coupling and optical excitation to off-resonant hyperfine states.

\begin{figure}
\centering{\includegraphics[width=0.7 \columnwidth]{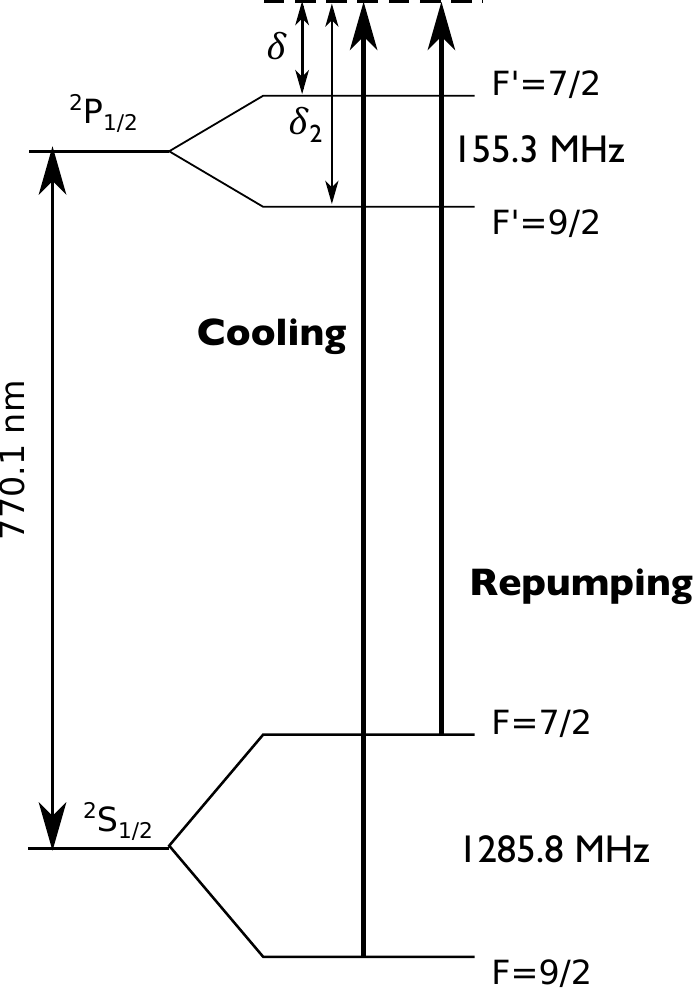}}
\caption{\label{fig:level scheme}Level scheme for the D$_1$ transition of $^{40}$K
and transitions used for gray molasses cooling. The laser detuning from the cooling/repumping transitions is $\delta$ and the detuning from the off-resonant excited hyperfine state $F'=9/2$ is $\delta_2$ (see text).}
\end{figure}

We implement 3D gray molasses cooling in $^{40}$K on the D$_{1}$ transition (see fig.~\ref{fig:level scheme}). In alkali atoms, the $P_{1/2}$ excited level manifold has only two hyperfine states, which are better resolved than their $P_{3/2}$ counterparts.
These facts allow for less off-resonant excitation and a good control of the cooling
mechanism. A first laser beam (cooling beam) is tuned to the $|^2 S_{1/2},F=9/2\rangle \to |^2 P_{1/2},F'=7/2\rangle$ transition with a detuning $\delta > 0$.  A second laser beam (repumping beam) is tuned to the $|^2 S_{1/2},F=7/2\rangle \to |^2 P_{1/2},F'=7/2\rangle $ transition with the same detuning $\delta$.

As mentioned above, two mechanisms can lead to the departure from the dark state. The first one is the motional coupling $V_{\rm mot}$ due to the spatial variations of the dark state internal wave-function induced by polarization and intensity gradients. The second one is the dipolar coupling $V_{\rm off}$ via off-resonant excited hyperfine states. A rough estimate shows that $V_{\rm mot}\simeq \hbar kv$, where $v$ is the velocity of the atom and $k$ the wave-vector of the cooling light, while  $V_{{\rm off}}\simeq\hbar\Gamma\left(\Gamma/\delta_{2}\right)I/I_{{\rm sat}}$, where $\Gamma^{-1}$ is the lifetime of the excited state, $I$ the light intensity, $I_{\rm sat}$ the saturation intensity and $\delta_2$ the detuning to off-resonant excited state. Comparing the two couplings, we see that the motional coupling is significant in the high velocity regime $v\gtrsim \Gamma/k\left(\Gamma/\delta_{2}\right)I/I_{{\rm sat}}$. In our case, the off-resonant level $F'=9/2$ (see fig.~\ref{fig:level scheme}) is detuned by $\delta_2=155.3\,\rm{MHz}+\delta$ from the cooling transition $|^2 S_{1/2},F=9/2\rangle \to |^2 P_{1/2},F'=7/2\rangle $. For $I\simeq I_{\rm sat}$, motional coupling dominates for $T\gtrsim 50\,\mu$K, meaning that both processes are expected to be present in our experiments. In general, the transition rate between $| \psi_{\rm{D}}\rangle $ and $| \psi_{\rm{B}}\rangle $ induced by motional coupling $V_{\rm{mot}}$ and the off-resonant coupling $V_{\rm{off}}$ are both maximal when the distance between the dark and bright manifolds is smallest, which favors transitions near the bottom of the wells of the optical lattice.

\begin{figure}
	\begin{subfigure}[b]{\columnwidth}
		\centering
		\includegraphics[width=0.9 \columnwidth]{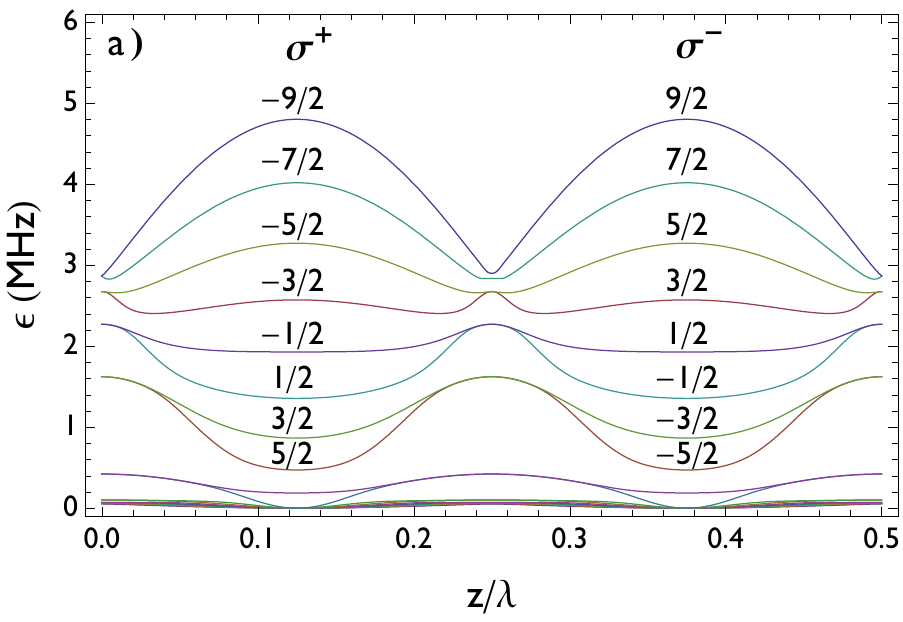}
	\end{subfigure}
	\begin{subfigure}[b]{\columnwidth}
		\centering
		\includegraphics[width=0.9 \columnwidth]{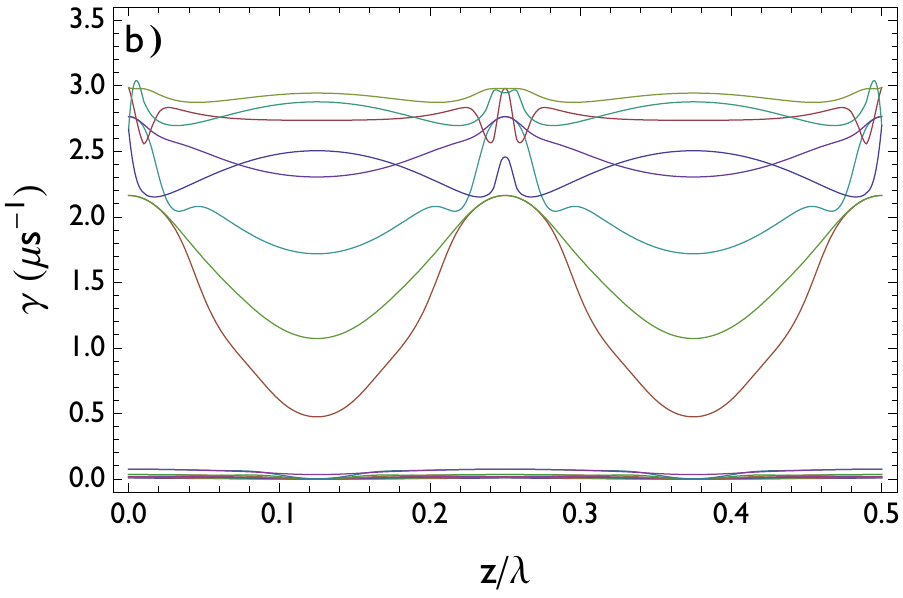}
	\end{subfigure}
	\begin{subfigure}[b]{\columnwidth}
		\centering
		\includegraphics[width=0.9 \columnwidth]{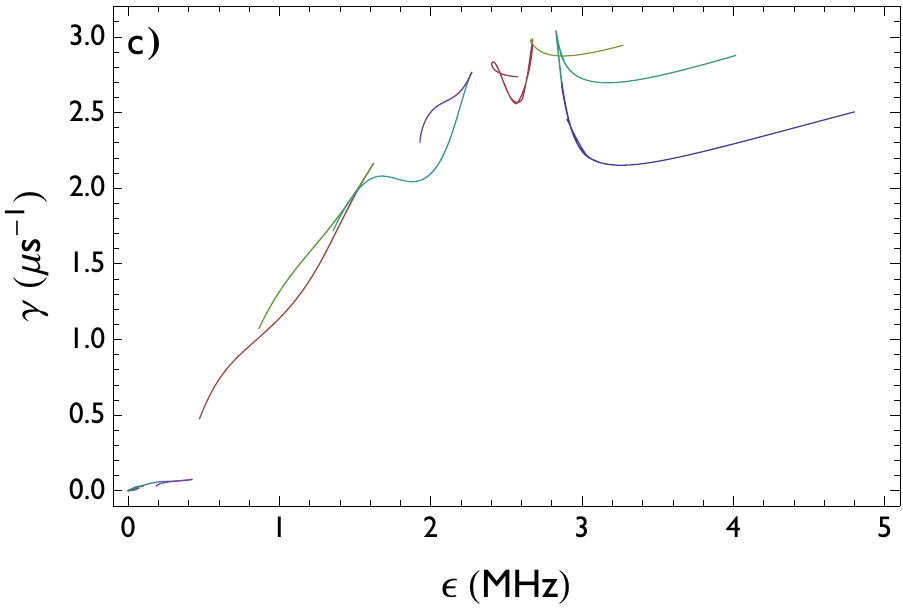}
	\end{subfigure}
	\caption{(Color online) Semi-classical calculation of the effect of dual frequency counter-propagating laser beams in a 1D lin$\perp$lin configuration on a $^{40}$K atom at rest. a) light shifts $\epsilon$ versus position, b) optical pumping rates $\gamma$. c) optical pumping rates versus light shifts. The laser intensities are $I_{\rm{cool}}=20I_{\rm{sat}}$ and $I_{\rm{repump}}=I_{\rm{cool}}/8$ per beam, with $\delta=+3\Gamma$. The different lines correspond to the $18$ dressed states of the $^2 S_{1/2}$ ground-state. At $z=\lambda /8$ the local polarization is $\sigma^+$ and here each curve corresponds to a pure $m_F$ state. At this position the light shift increases with $-m_F$. The $|^2S_{1/2},F=7/2\rangle $ manifold interacts only weakly with light since the repumping beam is kept at low intensity. Consequently, the light shifts and optical pumping rates are small.}\label{Fig:Saijun_figures}
\end{figure}

In $^{40}$K, the simplified discussion presented so far must be generalized to the case involving many hyperfine states ($10+8$). However, the essential picture remains valid. Indeed, by numerically solving the optical Bloch equations for the $^{40}$K system in the presence of the cooling and repumping laser fields, we obtain the light shifts $\epsilon$ and the optical pumping rates $\gamma$ of all the dressed states for an atom at rest (see fig.~\ref{Fig:Saijun_figures}). This is done for the particular case of a one-dimensional optical lattice in the lin$\perp$lin configuration and with a low repumping intensity (1/8 of the cooling beam intensity, typical for our experiments). In fig.~\ref{Fig:Saijun_figures}a) we see $8$ bright states, 2 weakly coupled states and 8 dark states combining both hyperfine manifolds. In fig.~\ref{Fig:Saijun_figures}b) we plot the optical pumping rates of the corresponding dressed states. We find that the optical pumping rate is low for the weakly coupled states and it practically vanishes for the dark states. In fig.~\ref{Fig:Saijun_figures}c) the optical pumping rates display a good correlation with the light shift magnitude, which favors efficient sub-Doppler cooling. Note also the long-lived dark states. This correlation shows that the gray molasses picture remains valid for this more complex level scheme.

We now turn to the question of the capture velocity of the gray molasses scheme.
Let $\Gamma'$ be the optical pumping rate from bright to dark states. The atom is pumped efficiently towards dark states if it stays a time $\tau\gtrsim \Gamma'^{-1}$ near the top of the hill. If the atom moves at a velocity $v$ in the lattice, then $\tau\simeq 1/kv$ and the optical pumping to dark states is efficient when $kv\lesssim \Gamma'$. $v_c\simeq \Gamma'/k$ thus defines the capture velocity of the gray molasses. For a beam with detuning $\delta$ to the main cooling transition, $\Gamma'\propto I/\delta^2$ and thus $v_c$ increases with laser intensity. On the other hand, the cooling efficiency is reduced when the atom cannot climb the potential hill anymore, which leads to an equilibrium temperature that scales as $k_B T\propto I/\delta$, when $T\gg T_{\rm Recoil}=\hbar^2 k^2 / 2 m k_{\rm B}$~\cite{Dalibard1989,Castin1991}.

\section{Experimental results}
Our setup is based on the apparatus presented in \cite{Ridinger2011}.
In the experiments presented here, $6.5\times 10^8$ $^{40}$K atoms are loaded from a two-dimensional magneto-optical trap (2D-MOT) into a three-dimensional magneto-optical trap (MOT) operating on the D$_2$ line. The initial temperature of the cloud is
$200\,\text{\ensuremath{\mu}K}$, not far from the Doppler temperature
$T_{D}=\hbar \Gamma/2 k_{\rm B} = 145\,\text{\ensuremath{\mu}K}$, with $\Gamma/2\pi\approx6.035$~MHz.
In the MOT, the cooling and repumping laser intensities are $I_{\rm cool}=13I_{\rm sat}$ and $I_{\rm repump}=I_{\rm cool}/20$
per beam, with $I_{\rm sat}=1.75$~mW/cm$^2$. After
the loading phase, we ramp the magnetic field gradient  from $9\,\text{G$\cdot$cm\ensuremath{^{-1}}}$
to $60\,\text{G$\cdot$cm\ensuremath{^{-1}}}$ in 5 ms without changing the laser detunings in order to compress the cloud. This process yields a cloud with high density , but with a much higher temperature of $\sim4\,\rm{mK}$. At
this point the magnetic field is switched off in  $\simeq100\,\mu s$
and the D$_{1}$ molasses beams are switched on for a time $\tau_{\rm m}$.

The D$_{1}$ cooling and repumping beams are detuned by the same amount $\delta$ in the range of $2\Gamma - 5\Gamma$ as  shown in fig.~\ref{fig:level scheme}. The repumping beam is detuned from the main cooling beam by $1285.8\,\text{MHz}$ using an electro-optical modulator. Its intensity is typically $1/8$ of the cooling beam intensity. After propagation through an optical fiber, the total D$_{1}$ optical power is $240\,\text{mW}$ and the beam is magnified to a waist of $1.1\,\text{cm}$. We then split the beam into two vertical
beams and two retro-reflected horizontal beams in a three-dimensional
$\sigma^{+}/\sigma^{-}$ configuration. The maximum D$_1$ cooling intensity per beam attained in our experiments is $25\,\text{mW/cm}^2$ or $I=14I_{\rm sat}$.

\begin{figure}
	\begin{subfigure}[b]{\columnwidth}
		\centering{\includegraphics[width=\columnwidth]{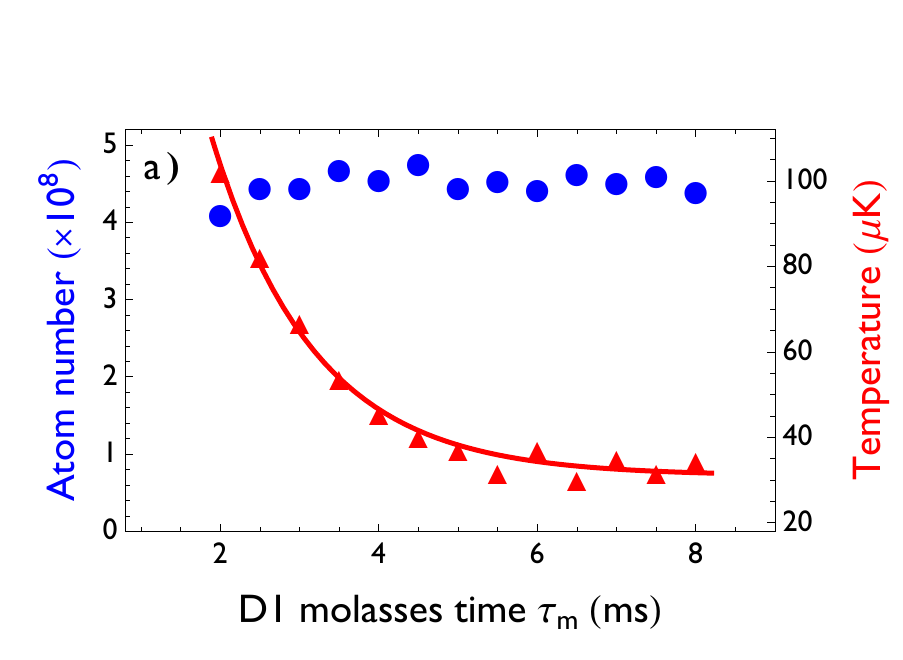}}
	\end{subfigure}
	\begin{subfigure}[b]{\columnwidth}
		\centering{\includegraphics[width=\columnwidth]{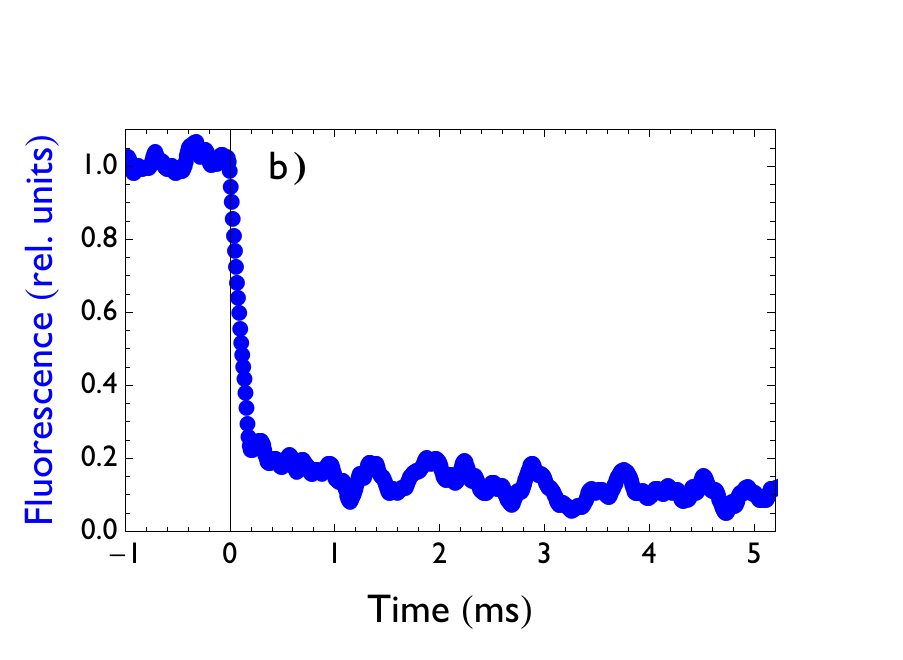}}
	\end{subfigure}

\caption{\label{fig:Dynamics}
(Color online) a) Number of atoms captured in the D$_1$ molasses (circles) and their temperature (triangles) as a function of molasses duration. The number of atoms in the compressed MOT was $4.5\times 10^8$. b) Measured fluorescence during the MOT and the D$_1$ molasses phase. Both experiments were performed with $I_{\rm cool}=14I_{\rm sat}$, $\delta=2.3\Gamma$ and $I_{\rm repump}=I_{\rm cool}/8$. }
\end{figure}

We first measure the atom number and temperature of the D$_{1}$ molasses as a function of the cooling beam duration $\tau_{\rm m}$ (fig.~\ref{fig:Dynamics}). The temperature is determined by time-of-flight. At high intensity $I_{\rm cool}=14I_{\rm sat}$ and detuning $\delta=2.3\Gamma$, all $4.5\times 10^8$ compressed MOT atoms are cooled to a temperature of $30\,\mu$K in 6 to 8 ms. Although the initial temperature of the compressed MOT is rather high, D$_{1}$ cooling occurs rapidly. As shown in fig.~\ref{fig:Dynamics}a), the temperature drops from $\sim 4\,$mK to 100~$\mu$K in 2~ms, and reaches its asymptotic value in about 6~ms. These dynamics are confirmed by direct measurement of the fluorescence light emitted during the D$_1$ molasses phase, as displayed in fig.~\ref{fig:Dynamics}b). The fluorescence exhibits a fast decay in $\sim 200\,\mu$s to about $20\%$ of the MOT light followed by a slower decay in $\sim 3\,$ms to $10\%$, which indicates the accumulation of atoms in weakly coupled states.

\begin{figure}
\begin{centering}
\includegraphics[width=\columnwidth]{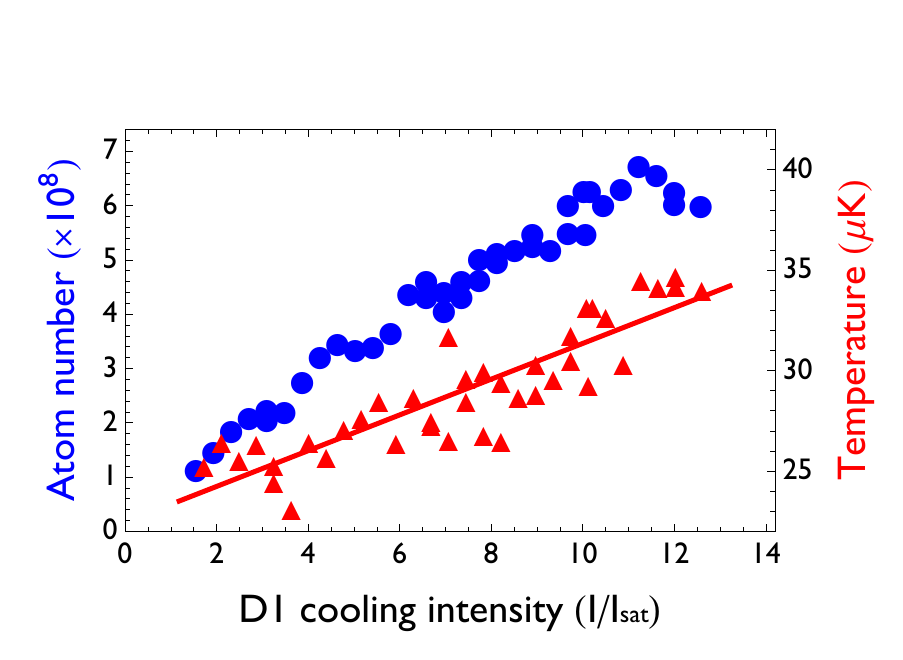}
\par\end{centering}

\caption{\label{fig:capture efficiency and temperature as function of intensity}(Color online) Number of atoms captured in the D$_1$ molasses (circles) and their temperature (triangles) as a function
of the $D_{1}$ cooling beam intensity for $\delta=2.3\Gamma$ and $I_{\rm repump}= I_{\rm cool}/8$. The number of atoms in the compressed MOT was $6.5\times 10^8$ and the capture efficiency reaches $\sim~100\%$ for $I\geq 11I_{\rm sat}$}
\end{figure}

When repeating the experiment for lower D$_{1}$ laser intensities for a fixed time of 6 ms, we observe both a decrease of the number of atoms cooled by gray molasses and a further lowering of the temperature down to $24\,\mu$K (fig.~\ref{fig:capture efficiency and temperature as function of intensity}). The number of atoms is measured after a time of flight of $20$~ms, after which we would not detect any atoms in the absence of D$_1$ molasses. The capture efficiency increases with the cooling intensity indicating a higher capture velocity at higher laser intensity and it reaches $\sim\,100\%$ for $I\geq 11I_{\rm sat}$. Similarly, the equilibrium temperature increases with laser intensity in the explored range in agreement with Sisyphus-type cooling mechanisms.

\begin{figure}
\begin{center}
\includegraphics[width=\columnwidth]{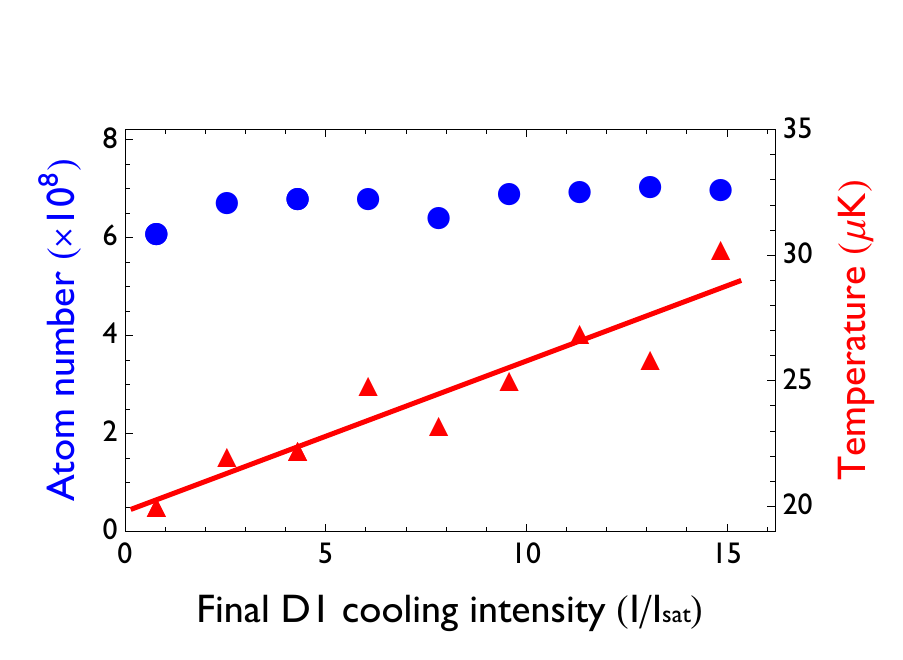}
\end{center}
\caption{(Color online) Number of atoms captured in the D$_1$ molasses (circles) and their temperature (triangles) after a 6~ms capture phase at high intensity $I_{\rm cool}=14I_{\rm sat}$ followed by a 2~ms linear intensity ramp to adjustable value. The detuning is fixed to $\delta=2.3\Gamma$. The number of atoms in the compressed MOT was $7\times 10^8$.}
\label{Fig:RampEfficiency}
\end{figure}

The results of fig.~\ref{fig:Dynamics}  and fig.~\ref{fig:capture efficiency and temperature as function of intensity} suggest implementing a cooling sequence with two successive phases. A first phase lasting 6-ms at high D$_1$ cooling intensity takes advantage of the high capture velocity. This phase is followed by a 2-ms stage in which the intensity is linearly reduced by an adjustable amount to further lower the temperature. As illustrated in fig.~\ref{Fig:RampEfficiency}, this supplementary cooling phase allows the sample to reach a temperature of $20\,\mu K$ by reducing the intensity by one order of magnitude and without any atom loss. No significant change of the atomic cloud volume was observed during this 8-ms sequence.

\begin{figure}
\begin{center}
\includegraphics[width=\columnwidth]{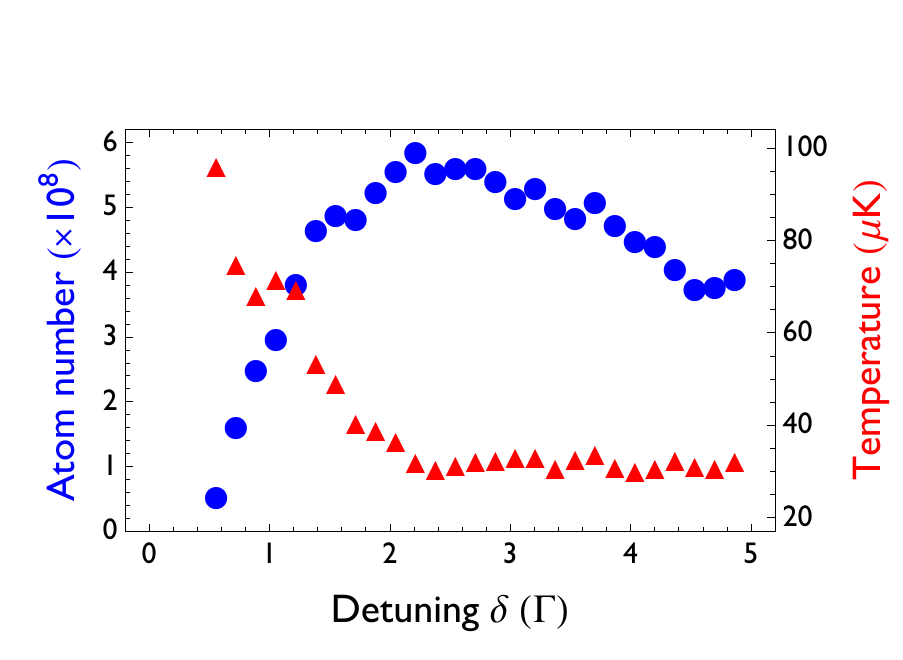}
\end{center}
\caption{(Color online) Number of atoms captured in the D$_1$ molasses (circles) and their temperature (triangles) for the dynamic 8~ms cooling sequence as a function of the detuning $\delta$. The number of atoms in the compressed MOT was $6\times 10^8$.}
\label{Fig:Detuning}
\end{figure}

In fig.~\ref{Fig:Detuning},  we show the number of atoms captured in the D$_1$ molasses and their temperature as a function of the laser detuning $\delta$ for the complete 8~ms sequence. For $\delta \in [0.5\Gamma, 2\Gamma]$, we observe a steep decrease of the temperature from $100\,\mu$K to $30\,\mu$K, as expected from Sisyphus cooling, followed by a plateau near $30\,\mu$K for detunings above $2\Gamma$. The capture efficiency raises sharply to $\sim 100\%$ at $\delta \sim 2.3 \Gamma$, displays a broad maximum and slowly decreases above $4\Gamma$, indicating a decrease of the capture velocity. Finally, optimal parameters for $^{40}$K gray molasses are summarized in Table \ref{Table:OptimalParameters}.

\begin{table}[!]
\begin{center}
\begin{tabular}{c|ccc}
&duration (ms)&$I_{\rm cool} (I_{\rm sat})$&$\delta (\Gamma)$\\
\hline
\hline
capture phase&6&14&+2.3\\
cooling phase&2&14$\to$1&+2.3\\
\hline
\end{tabular}
\end{center}
\caption{Optimized parameters for $^{40}$K D$_1$ gray molasses. Using these parameters, all the $6.5\times 10^8$ atoms from a compressed MOT are cooled to $20\,\mu$K in D$_1$ gray molasses.}
\label{Table:OptimalParameters}
\end{table}

We checked that the minimum temperature of $20\, \mu$K is not limited by residual magnetic fields nor by atomic density. We found that the residual magnetic field during the D$_1$ molasses should be minimized. Indeed, introducing a small tunable bias magnetic field B in the vertical direction, the D$_1$ molasses temperature increased quadratically as $\Delta T\approx\,80\rm{B}^{2}\text{\ensuremath{\mu}K/G\ensuremath{^{2}}}$. For this reason, the stray magnetic field was cancelled to less than 100 mG in three directions using compensation coils. We also searched for a density dependent temperature limitation and observed no significant temperature change when the density was reduced by a factor of 4 from $n_0 \sim 2\times 10^{10} \rm{cm}^{-3}$. Modeling gray molasses cooling in three dimensions in order to understand the temperature limit remains today an open problem.

\section{Magnetic trapping}
After the D$_1$ molasses phase, the atoms are optically pumped to the $|^2S_{1/2},F=9/2,m_{F}=9/2\rangle$ stretched state and then transferred into a quadrupole magnetic trap. The $\sigma^+$ polarized optical pumping laser beams are pulsed for $120\,\mu$s in the presence of a bias magnetic field. After this phase, the trap axial magnetic field gradient is raised from 0 to 37 G$\cdot$cm$^{-1}$ in 3 ms, followed by a compression to 76 G$\cdot$cm$^{-1}$ in 147 ms and a thermalization stage lasting 350 ms during which the field gradient remains constant. At this point we detect $2.5\times10^{8}$ atoms at a temperature of
$80\,\mu K$. Assuming that all atoms are in the $|F=9/2,m_F=9/2\rangle$ stretched state, the central phase-space density is $\text{PSD}=n_0 \lambda^3_{\rm dB}\approx 2\times10^{-5}$. From the {\it p-wave} cross-section $\sigma \approx 2\times 10^{-11}\,\rm{cm}^2$ at a temperature of $80\,\mu$K measured in~\cite{DeMarco1999}, we estimate the trap averaged initial collision rate to be $\gamma_{\rm coll}=n_0 \sigma \bar{v}/8\sqrt{2}\approx 23\,\rm{s}^{-1}$. This rate is quite favorable for initiating evaporative cooling.

\section{Conclusion}
We have shown that gray molasses operating on the D$_1$ optical transition is a very simple and powerful method to increase the  phase space density of laser-cooled $^{40}$K alkali gases to $\sim  10^{-5}$. This phase space density leads to excellent starting conditions for evaporative cooling in magnetic or optical dipole traps. For $^{40}$K, this is particularly useful as the low temperature allows direct transfer into an optical trap and magnetic tuning to a Feshbach resonance for efficient evaporation. Moreover, our results open the way for sub-Doppler cooling of other
atoms with narrow $P_{3/2}$ excited states, such as $^{6}$Li and  $^{7}$Li.  We already have experimental evidence for sub-Doppler D$_1$ cooling of $^{6}$Li and $^{7}$Li and this will be the subject of a future publication.

\acknowledgments
We acknowledge useful discussions with J. V. Porto, J. Dalibard, L. Khaykovich and D. Suchet.
We acknowledge support from R\'egion Ile de France (IFRAF), EU (ERC
advanced grant Ferlodim) and Institut Universitaire de France. D.R.F.
acknowledges the support of Funda\c{c}\~{a}o para a Ci\^{e}ncia e Tecnologia (FCT-Portugal), through the grant
number SFRH/BD/68488/2010.

\bibliographystyle{eplbib.bst}
\bibliography{D1Cooling}
\end{document}